# Robust integer and fractional helical modes in the quantum Hall effect

*Yuval Ronen[†], Yonatan Cohen[†], Daniel Banitt, Moty Heiblum[#] and Vladimir Umansky*

*Braun Center for Submicron Research, Department of Condensed Matter Physics,*

*Weizmann Institute of Science, Rehovot 76100, Israel*

[†] *equal contributions*

[#] *corresponding author (moty.heiblum@weizmann.ac.il)*

**Electronic systems harboring one-dimensional helical modes, where the electron's spin and momentum are locked, have lately become an important field of its own. When coupled to a conventional superconductor, such systems are expected to manifest topological superconductivity; a unique phase that gives rise to exotic Majorana zero modes. Even more interesting are fractional helical states which have not been observed before and which open the route for the realization of the generalized para-fermions quasiparticles. Possessing non-abelian exchange statistics, these quasiparticles may serve as building blocks in topological quantum computing. Here, we present a new approach to form protected one-dimensional helical and fractional helical edge modes in the quantum Hall regime. The novel platform is based on a carefully designed double-quantum-well structure in a high mobility GaAs based system. In turn, the quantum well hosts two sub-bands of 2D electrons; each tuned to the quantum Hall effect regime. By electrostatic gating of different areas of the structure, counter-propagating integer, as well as fractional, edge modes (belonging to Landau-levels with opposite spins) are formed – rendering the modes helical. We demonstrate that due to spin protection, these helical modes remain ballistic, without observed mixing for large distances. In addition to formation of helical modes, this new platform can be exploited as a rich playground for an artificial induction of compounded fractional edge modes, as well as construction of interferometers based on chiral edge modes.**

Pursuing Majorana zero modes (MZMs) in condensed matter physics is gaining wide range interest[1–9]. While bearing some resemblance to their high energy counterparts, condensed matter MZMs are significantly different as they are expected to possess non-abelian exchange statistics, which renders them as potential candidates for topologically protected qubits[2,10–15]. One of the most promising platforms for the formation of MZMs is a one-dimensional helical system coupled to an s-wave superconductor[16,17]. In a helical system, electrons moving in opposite directions possess opposite spins (spin degeneracy is lifted), while the two original spin species are still present. Their coupling to an s-wave superconductor, induces topological, 'spinless', p-wave paring. At the two ends of the induced superconductor, two localized MZMs are expected to form. Even more interesting are the generalized para-fermions, which are expected to emerge when coupling a conventional superconductor to helical modes in the fractional quantum Hall effect regime[18–20].

Most of the present attempts to form helical modes concentrate in materials with strong spin-orbit coupling[5–9,21,22]. While signatures of localized MZMs appeared, the helical nature of the underlying modes is not confirmed. Lately, the presence of helical edge modes was reported in small size (~350nm) 2D topological insulators[21–23], as well as in a twisted bilayer graphene in the IQHE regime (of size ~15µm)[24]. Another work attempted to form helical modes by doping of magnetic impurities in quantum wells and electrostatically inducing ferromagnetic transitions[25]. However, in those works spin protection from backscattering was not reported, while strong inter-mode mixing limited the propagation length. Moreover, the formation of fractional helical modes was not reported yet.

We developed a new platform that enables formation of robust and highly controllable helical modes in the QHE effect regime. The platform is based on a 2-dimensional electron gas (2DEG) embedded in a double-quantum-well (DQW), which hosts two electronic sub-bands. By a proper electrostatic gating of two adjacent half-planes of the 2DEG, spin-split Landau levels (LLs) belonging to the different sub-bands cross each other at the interface between the half-planes forming counter propagating edge modes. When the intersecting LLs possess opposite spins, the modes are therefore helical.

Figure 1 illustrates schematically the formation of helical edge modes in the two sub-bands quantum Hall system. The sub-bands, denoted by $L_1$ and $L_2$, are depicted as two 2-dimentional sheets (blue for $L_1$ and red for $L_2$). Each sub-band splits to discrete LLs at high magnetic field, with individual filling factors, $\nu_1$ in $L_1$ and $\nu_2$ in $L_2$; with a generalized filling factor $\nu = (\nu_1, \nu_2)$. Figures 1a & 1b describe the scenarios of $\nu = (2,0)$ and $\nu = (1,1)$, respectively. When these two configurations are placed near each other, as shown in Fig. 1c, counter-propagating chiral edge modes, with opposite spins, propagate along the interface (spin-down in $L_1$, blue, and spin-up in $L_2$, red), manifesting integer helical edge modes.

Figure 2a shows a schematic illustration of the heterostructure used to implement the two sub-bands system. A 40nm thick GaAs layer, cladded on top and bottom by AlGaAs layers, forms the quantum well structure. A thin AlAs barrier, 3nm thick, is inserted in the middle of the GaAs layer to form a DQW potential landscape. The total areal density $n = 2.7 \times 10^{11} cm^{-2}$ and the low temperature mobility $\mu = 0.6 \times 10^6 cm^2 V^{-1} s^{-1}$. Modulation doping, predominantly at the lower side of the DQW, leads to a tilted potential in the well (self-consistent simulation in Fig. 2b). An SEM image in Fig. 2c shows a fabricated structure with its top-gates, splitting the surface to two half-planes (see Methods).

The fundamental difficulty in realizing the proposed configuration is illustrated schematically Fig. 2d, where a naïve illustration of the energy dependence of the LLs (in $L_1$ and in $L_2$) is plotted as function of magnetic field. The two different generalized fillings; *e.g.*, (2,0) and (1,1), each in a different half-plane, cannot coexist in a single magnetic field. However, in actual practice the situation is different[26]. As the energy of different LLs (of different sub-bands) cross, charge must redistribute between the sub-bands. The charge redistribution leads to bending of the linear-like evolution of the LLs' fan-diagram, allowing, under a proper design, for two generalized fillings (with equal sums of the individual fillings on both sides) to take place at the same magnetic field. Note, that charge transfer costs an additional energy since it charges the capacitance between the two regions of the DQW, thus partly opposing charge transfer[26]. A narrower DQW is desirable for a greater charge transfer, and thus a more pronounced bending away from the linear evolution of the LLs with magnetic field.

The fan-diagram of the longitudinal resistance (which follows the evolution of the LLs) is plotted in Fig. 2e. As the top gate voltage rises above -0.15V, LLs in $L_2$ gradually fill and charge transfers from $L_1$ to $L_2$ (as LLs cross). LLs in $L_1$ lose carriers, thus shifting to lower magnetic fields ($L_1$ LLs' lines have negative slopes around the crossing regions). In the present configuration, there are multiple filling fractions between *ν*=2 and *ν*=1; hence, the charge transfer near the crossings of LLs is not large enough to allow a gate-controlled transition $(2,0)\rightarrow(1,1)$. Yet, a gate-controlled transition, $(4,0)\rightarrow(3,1)$, can be achieved (Fig. 3a). With $B=2.275T$ and $V_{LG}=-0.158V$, the general filling factor at the left half-plane is $v=(4,0)$, while $V_{RG}$ scanning along the black arrow varies the filling on the right from $v=(3,0)$, to $v=(4,0)$, and finally to $v=(3,1)$. The resultant configuration is shown in Fig. 3b.

A current of 1nA is injected at $S_1$ and its reflected part, $I_{S1\rightarrow D1}$, is plotted in Fig. 3c (upper panel). When the right half-plane is tuned to $v=(3,0)$ or $v=(4,0)$, all injected edge modes circulate the outer perimeter of the 2D plane, arriving at $D_2$, while $I_{S1\rightarrow D1}=0$. Yet, when the right half-plane is tuned to $v=(3,1)$, three edge modes are transmitted to $D_2$, while one (the lowest LL of $L_2$) flows across the interface of the half-planes and is fully reflected to $D_1$, leading to $I_{S1\rightarrow D1}=0.25nA$. Similarly, injecting current at $S_2$ and measuring the reflected current in $D_2$, $I_{S2\rightarrow D2}$, leads to complimentary results (Fig. 3c, lower panel). These observations clearly agree with ballistic propagation of helical modes (Fig. 3c, green shaded region).

The $(4,0)\rightarrow(3,1)$ transition is evidently only one example where interlayer charge transfer is sufficiently large to allow formation of helical modes. Figure 2e shows four such transitions $(n,0)\rightarrow(n-1,1)$ with $n$=3, 4, 5, 6, which allow gate-controlled transitions (open circles). Transitions with an even $n$, such as $(4,0)\rightarrow(3,1)$ and $(6,0)\rightarrow(5,1)$ (green circles), lead to helical modes. Same spins counter-propagating modes are born with the transitions $(3,0)\rightarrow(2,1)$ or $(5,0)\rightarrow(4,1)$ (red circles).

In Fig. 4 measurement results of $I_{S1\rightarrow D1}$ are plotted for the four transitions mentioned above in devices with three different counter-propagation lengths, $L_{CP}$=7μm, 150μm, and 300μm. A clear difference is observed between same-spin transitions $(odd,0)\rightarrow(even,1)$ and opposite-spin $(even,0)\rightarrow(odd,1)$ transitions. In the former case, as $L_{CP}$ increases beyond 7μm a reduction in $I_{S1\rightarrow D1}$ corresponding to an approximate equilibration length of ~1mm, is observed. Evidently, a reduction in $I_{S1\rightarrow D1}$ is compensated by an increase in the transmitted current $I_{S1\rightarrow D2}$ (see SI, section S1), indicating tunneling between the edge modes (with no bulk current). In contrast, when helical modes are formed, no reduction in $I_{S1\to D1}$ is observed even for $L_{CP}$=300μm - demonstrating spin protection.

We turn to the fractional regime and concentrate on $R_{XX}$ in the $B$-$V_{RG}$ plane around the $(2,0)\to(1,1)$ transition (Fig. 5a). The red and yellow dots, which stand for $(4/3,0)$ and $(1,1/3)$, allow an intersection of counter-propagating edge modes with opposite spins; each with filling $v=1/3$ (Fig. 5b). Indeed in the appropriate $V_{RG}$ range, the reflected currents $I_{S1\to D1}$ & $I_{S2\to D2}$ are found to be $I_{S1\to D1}=0.25nA$ & $I_{S2\to D2}=0.25nA$, respectively (Fig. 5c, green region). Note, that while $I_{S2\to D2}$ is not affected by the propagation length, $I_{S1\to D1}$ decreases slightly as the propagation length increases; however, that decrease is not accompanied by an increase in the transmitted current, $I_{S1\to D2}$ (see SI, section S2). Indeed, the missing current flows through the bulk in the right half-plane due to its finite $R_{XX}$ (SI, S2).

Finally, by directly contacting the helical modes, establishing thus a common Fermi energy in the two counter-propagating modes (Fig. 6a), four-terminal measurements can be performed (Fig. 6b). Current $I$ is injected in contact #3 while contacts #1, #2 and #6 are grounded. The ratio between the potential difference between contacts #4 and #5, $V_{45}$, and the current $I$ is the appropriate trans-resistance. The voltage $V_{45}$ is plotted as function of the magnetic field in Fig. 6d, for $V_{LG}=-0.158V$ and $V_{RG}=-0.09V$ (denoted by the white dashed lines in Fig. 6c). At low and high magnetic fields, with the transitions $(4,0)\to(4,1)$ and $(3,0)\to(3,1)$, only a single chiral channel carries the current along the interface between the two regions; hence, $V_{45}$=0. However, in the helical regime, with the transition $(4,0)\to(3,1)$, two counter-propagating edge modes carry the currents between the contacts and $V_{45}/I=\frac{R_Q}{4}$, where $R_Q=\frac{h}{e^2}$; in a good agreement with the expected trans-resistance.

2DEG embedded in GaAs-AlGaAs heterostructures did not play any significant role thus far in the emergent field of topological insulators and superconductors (aside, of course, from the illustrious QHE). This is a direct result of the very weak spin-orbit coupling and the difficulties in inducing superconductivity in the buried 2DEG. Yet, the advantage of high mobility electrons, the ease in processing complex structures, and the well-established robust QHE states (integer and fractional), make this material system highly attractive. Here, by employing a DQW in the integer and fractional QHE regime, robust and strongly protected ballistic helical modes are formed. Moreover, the spin protection provided by the helical modes is shown to increases the ballistic propagation length significantly.

Aside from the obvious next step of inducing superconductivity in the 2D electrons, and thus forming MZMs or para-fermions, this versatile implementation leads itself also to host non-abelian quasiparticles in topological defects, which do not require induced superconductivity[27]. Moreover, this new platform can serve as a versatile playground for investigating compounded QH edge modes and their mutual interaction. For example, the spontaneous emergent of counter-propagating QH edge modes, such as hole-conjugate states (*e.g.*, $v$=2/3, polarized and unpolarized), can be artificially created by intersecting $v$=1 and $v$=-1/3 states in a highly controlled fashion, allowing thus testing the transition from the never observed upstream current modes to upstream neutral modes[28–30].

## Methods

### Sample fabrication

An etch-defined Hall-bar with NiGeAu ohmic contacts fabricated using E-beam lithography. This followed by an atomic layer deposition of $HfO_2$, E-beam lithography, E-gun evaporation of 5/20nm Ti/Au top gates. The top gates, each defines a half-plane of the 2DEG, are separated by a gap of 80nm. Finally, the $HfO_2$ is etched in small regions of the contacts, connected to the bonding pads by 5/120nm Ti/Au leads. connected to bonding pads were evaporated.

**Acknowledgements:**

We acknowledge Johannes Nübler, Erez Berg, Yuval Oreg, Ady Stern, Yuval Gefen, Jinhong Park, Dmitri Feldman and Kyrylo Snizhko for fruitful discussions. We thank Diana Mahalu in the Ebeam processing and Vitaly Hanin for the help in the ALD process. M.H. acknowledges the partial support of the Israeli Science Foundation (ISF), the Minerva foundation, the U.S.-Israel Bi-National Science Foundation (BSF), the European Research Council under the European Community's Seventh Framework Program (FP7/2007-2013)/ERC Grant agreement No. 339070, and the German Israeli Project Cooperation (DIP).

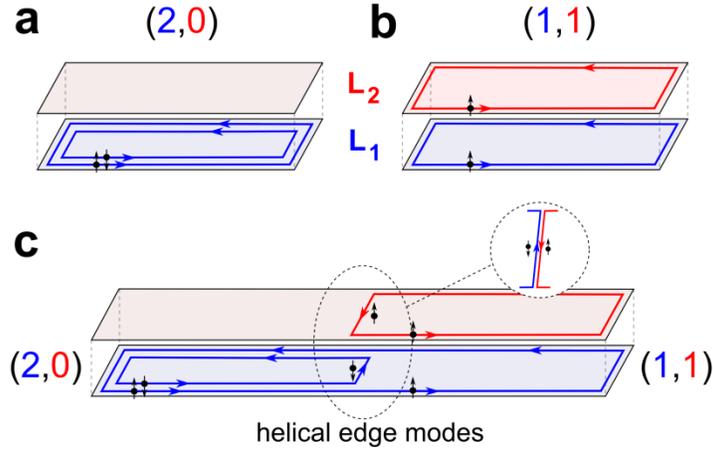

**Figure 1. Schematic illustration of the concept of creating helical edge modes in a double-layer quantum Hall effect system.** Two parallel 2DEGs layers are shown, one in blue denoted by $L_1$, and one in red denoted by $L_2$. Each layer has its own filling factor, $\nu_1$ and $\nu_2$, respectively. The double-layer generalized filling denoted is $\nu = (\nu_1, \nu_2)$. **a,** Scenario of $\nu = (2,0)$, with two edge modes propagating at the edge of $L_1$. **b,** Scenario of $\nu = (1,1)$, with one edge modes propagating at the edge of $L_1$ and one at the edge of $L_2$. **c,** The left half-plane is in $\nu = (2,0)$ and the right half-plane is in $\nu = (1,1)$. This creates counter propagating edge modes with opposite spins at the interface between the two half-planes (see inset).

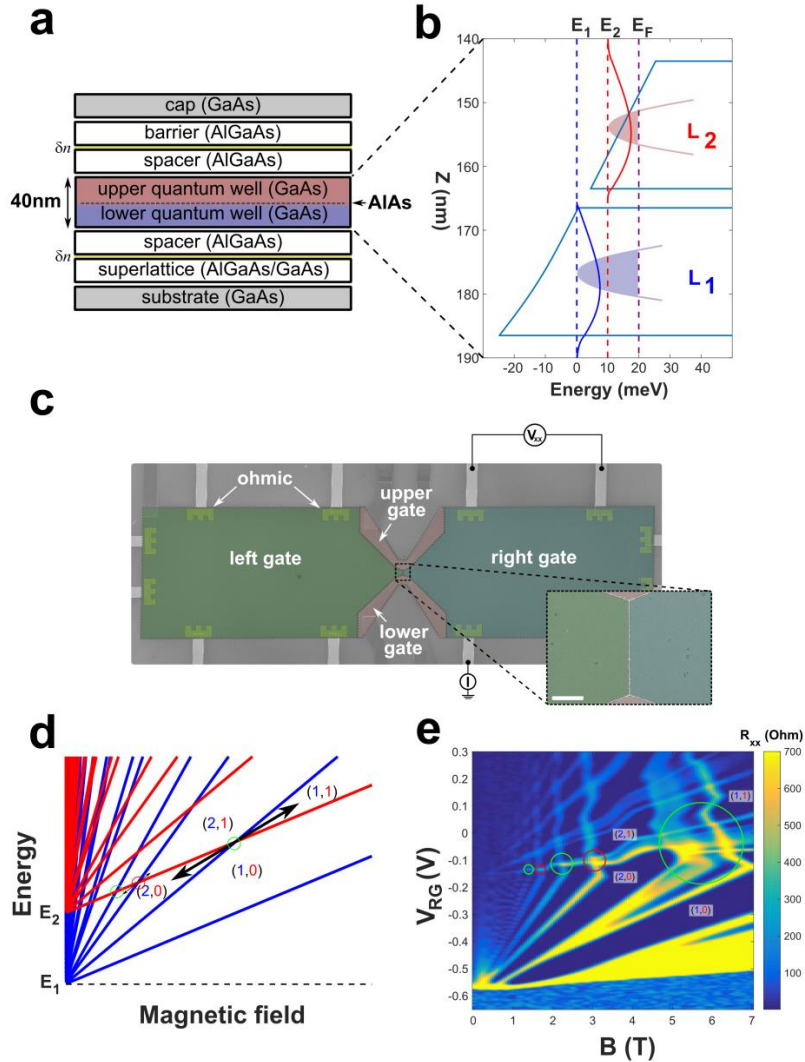

# Fig 2

**Figure 2. MBE growth sequence, lithographic patterning, and actual fan diagram. a,** schematic view of the double quantum well heterostructure. The lower and upper GaAs quantum wells are colored in blue and red, respectively. Each well is ~20nm wide with a 3nm AlAs barrier separating the two wells. **b,** Simulation of the potential landscape and of the 2D wave-functions at zero magnetic field. **c,** False colors SEM image of the device. Note, the four top gates allow changing configurations with the gates' voltages. Inset: zoom on the interface between the left and right top gates where the helical edge modes are designed to emerge. (scale-bar 2μm). **d,** Ideal energy fan diagram for the two-layer 2DEG. The energies of the LLs of $L_1$ (blue) and $L_2$ (red) are plotted as a function of magnetic field. **e,** Measurement of the longitudinal resistance, $R_{XX}$, of the right side of the device (Fig. 2c) as a function of magnetic field and gate voltage $V_{RG}$. The circles mark regions of LLs crossings with either opposite spins (green circles) or same spins (red circles).

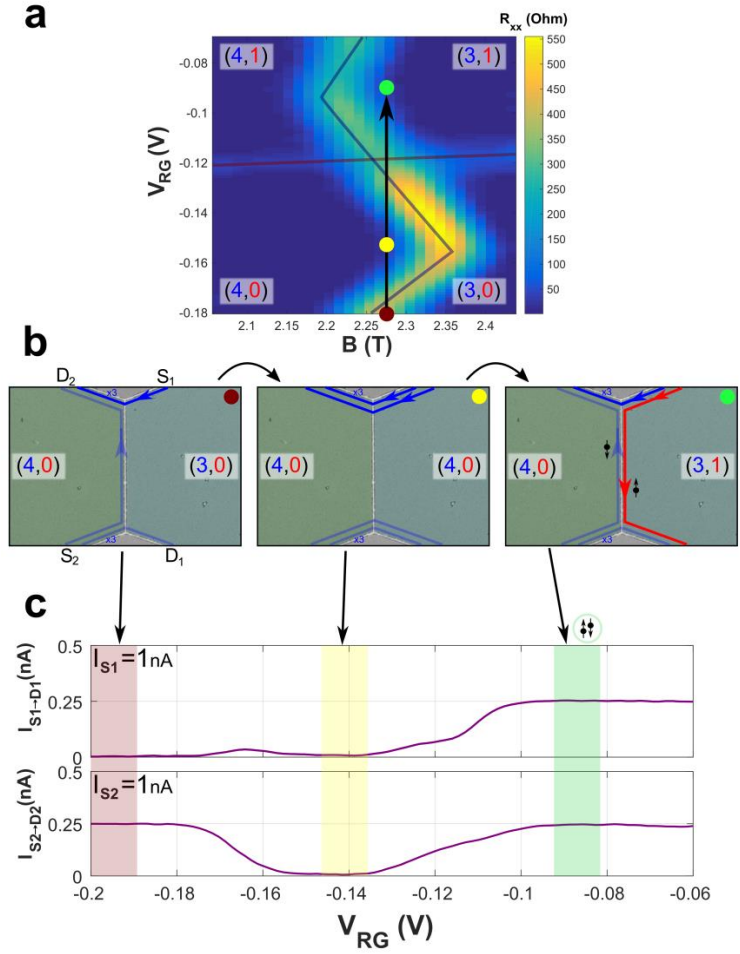

# Fig 3

**Figure 3. Tuning the structure to host integer helical modes. a,** Longitudinal resistance, $R_{XX}$, of the right side of the device as a function of magnetic field and gate voltage, $V_{RG}$, near the (4,0) - (3,1) transition. **b,** An illustration of the propagating edge modes when the left side is set to (4,0) and the right side is scanned along the black arrow, *i.e.* from (3,0) (red point) through (4,0) (yellow point) to (3,1) (green point). **c,** The top graph corresponds to the measured reflected current in $D_1$ when current of 1nA is injected from contact $S_1$, with the left side is held at (4,0) and the gate voltage on the right is scanned. The red, yellow and green shaded regions mark the gate voltage ranges leading to general fillings (3,0), (4,0) and (3,1) of the right side. The bottom figure shows the measurement of the reflected current reaching contact $D_2$ when a current of 1nA is injected from contact $S_2$.

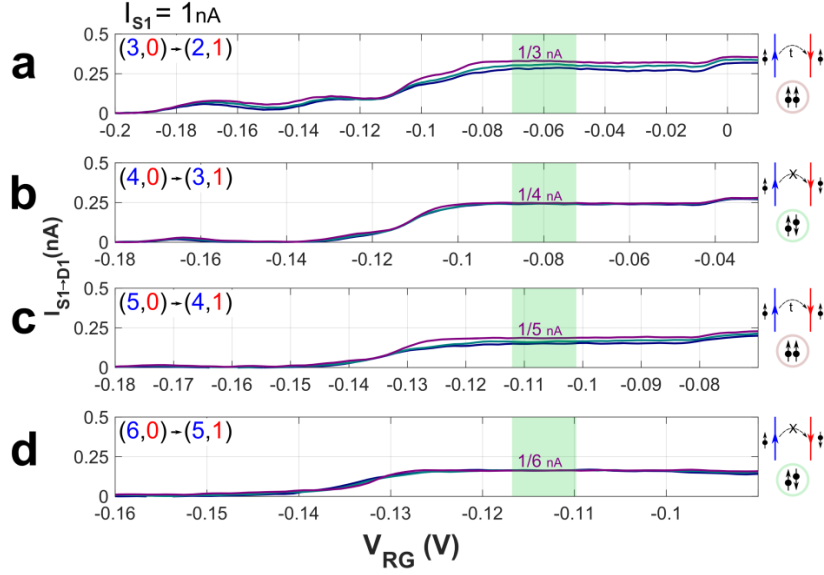

**Figure 4. Spin protected inter-mode tunneling. a-d,** Measurements of the reflected current reaching contact $D_1$ when current is injected from contact $S_1$ for different filling factors. In all the measurements the left side is held at $(n, 0)$ and the right side is scanned from $(n-1, 0)$, through $(n, 0)$ and to $(n-1, 1)$, where $n=3,4,5,6$ in **a**, **b**, **c** and **d**, respectively. The green shaded regions mark the voltage range in which counter-propagating intersecting edge modes are formed. The different colors, purple, green and blue, correspond to three propagation lengths, $L_{CP}=7, 150, 300\mu m$, respectively. The cases where $n$ is odd correspond to the two edge modes having the same spin orientation, while the cases where $n$ is even correspond to the two edge modes having opposite spins. Evidently, while inter-mode tunneling is evident for the same-spin configurations, transport remain ballistic, even for $L_{CP}=300\mu m$, in the opposite spin cases.

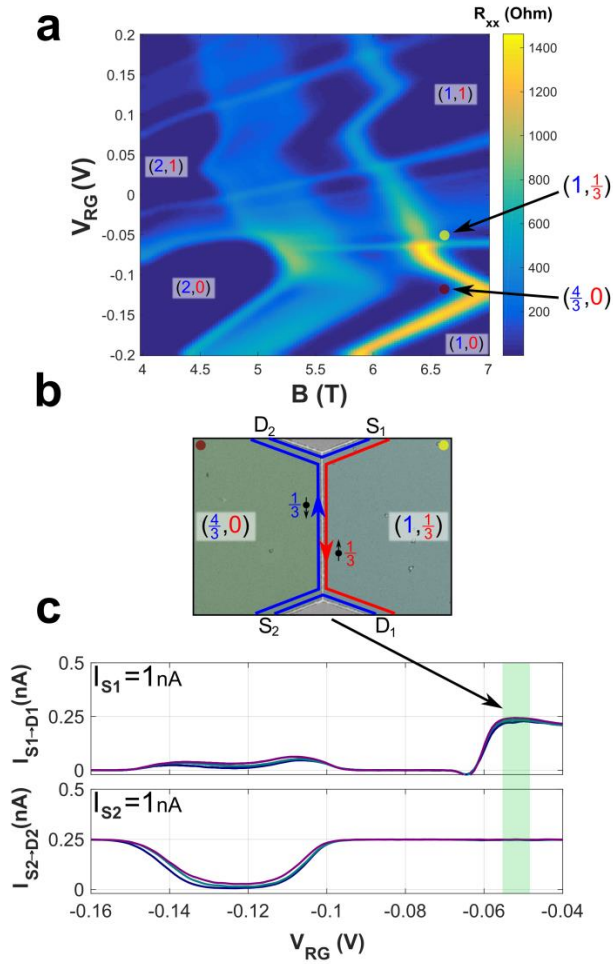

# Fig 5

**Figure 5. Formation of fractional helical states. a,** Fan diagram of the longitudinal resistance, $R_{XX}$, of the right side of the device as a function of magnetic field and gate voltage $V_{RG}$, in the region around the (2,0)-(1,1) transition. Fractional quantum Hall states are clearly observed within the regions of the integer states. The red and yellow dots correspond to filling factors (4/3,0) and (1,1/3), respectively, and emerge at the same magnetic field. **b,** Illustration of the formed fractional helical states. **c,** The top (bottom) graph displays the reflected current reaching $D_1$ ($D_2$), when current is injected from $S_1$ ($S_2$), while the left side is tuned to (4/3,0) and the gate voltage on the right is scanned. The shaded green region corresponds to the voltage range in which the right side is at (1,1/3). In both figures the reflected current in this voltage range is 0.25nA (fully reflected 1/3$^{rd}$ edge mode from both sides).

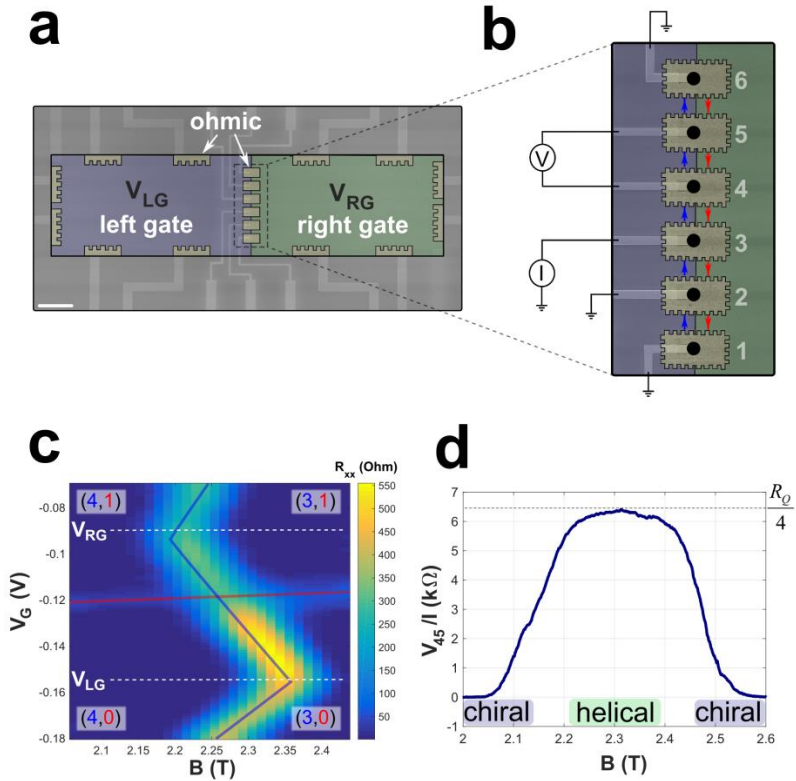

**Figure 6. Contacting directly the helical edge modes. a,** False colors SEM image of the device. Six ohmic contacts are alloyed at the interface between the top left gate (purple) and the top right gate (green). **b,** Zoom in on the region of the contacts with the measurement scheme. Current is injected at contact #3 and voltage is measured between contacts #4 and #5, $V_{45}$, while the other contacts are grounded. **c,** The longitudinal resistance, $R_{XX}$, of the right side as a function of magnetic field and gate voltage, $V_{RG}$, at the right side near the (4,0)-(3,1) transition. **d,** Evolution of the voltage $V_{45}$ as function of magnetic field, with $V_{RG}$ and $V_{LG}$ are fixed (white dashed lines in c). In the range of magnetic field where only a single chiral edge mode propagates along the interface, $V_{45}=0$. Yet, in the field range where helical edge modes form, $V_{45}/I=R_Q/4$ - as expected.